# Spatial mapping of protein composition and tissue organization: a primer for multiplexed antibody-based imaging

John W. Hickey[1,‡], Elizabeth K. Neumann[2,3,‡], Andrea J. Radtke[4,‡,*], Jeannie M. Camarillo[5], Rebecca T. Beuschel[4], Alexandre Albanese[6,7,¥], Elizabeth McDonough[8], Julia Hatler[9], Anne E. Wiblin[10], Jeremy Fisher[11], Josh Croteau[12], Eliza C. Small[13], Anup Sood[8], Richard M. Caprioli[2,3,14], R. Michael Angelo[1], Garry P. Nolan[1], Kwanghun Chung[6,7,15-18], Stephen M. Hewitt[19], Ronald N. Germain[4], Jeffrey M. Spraggins[3,14,20], Emma Lundberg[21], Michael P. Snyder[22], Neil L. Kelleher[5], Sinem K. Saka[23,24,*]

[1]Department of Pathology, Stanford University School of Medicine, Stanford, CA 94305, USA.
[2]Department of Biochemistry, Vanderbilt University, Nashville, TN, 37232 USA.
[3]Mass Spectrometry Research Center, Vanderbilt University, Nashville, TN 37232, USA.
[4]Lymphocyte Biology Section and Center for Advanced Tissue Imaging, Laboratory of Immune System Biology, NIAID, NIH, Bethesda, MD 20892, USA.
[5]Department of Chemistry, Molecular Biosciences and the National Resource for Translational and Developmental Proteomics, Northwestern University, Evanston, IL 60208, USA.
[6]Institute for Medical Engineering and Science, MIT, Cambridge, MA, USA.
[7]Picower Institute for Learning and Memory, MIT, Cambridge, MA, USA.
[8]GE Research, Niskayuna, NY,12309, USA.
[9]Antibody Development Department, Bio-techne, Minneapolis, MN 55413, USA.
[10]Department of Research and Development, Abcam PLC, Discovery Drive, Cambridge Biomedical Campus, Cambridge, CB2 0AX, UK.
[11]Department of Research and Development, Cell Signaling Technology, Inc., Danvers, MA 01923, USA.
[12]Department of Applications Science, BioLegend, San Diego, CA 92121, USA.
[13]Thermo Fisher Scientific, Rockford, IL, USA.
[14]Department of Chemistry, Vanderbilt University, Nashville, TN 37232, USA.
[15]Department of Chemical Engineering, MIT, Cambridge, MA, USA.
[16]Department of Brain and Cognitive Sciences, MIT, Cambridge, MA, USA.
[17]Center for Nanomedicine, Institute for Basic Science (IBS), Seoul, Republic of Korea.
[18]Yonsei-IBS Institute, Yonsei University, Seoul, Republic of Korea.
[19]Laboratory of Pathology, Center for Cancer Research, National Cancer Institute, National Institutes of Health, Bethesda MD 20892, USA.
[20]Department of Cell and Developmental Biology, Vanderbilt University School of Medicine, Nashville, TN 37240, USA.
[21]Science for Life Laboratory, School of Engineering Sciences in Chemistry, Biotechnology and Health, KTH–Royal Institute of Technology, Stockholm, Sweden.
[22]Department of Genetics, Stanford University School of Medicine, Stanford, California 94305, USA.
[23]Wyss Institute for Biologically Inspired Engineering at Harvard University, Boston, MA 02115, USA.
[24]European Molecular Biology Laboratory (EMBL), Genome Biology Unit, 69117 Heidelberg, Germany.

‡Authors contributed equally to this work and are listed alphabetically.
¥Current Address: Boston Children's Hospital, Division of Hematology/Oncology, Boston, MA, USA
*Corresponding Authors

**Abstract**

Tissues and organs are composed of distinct cell types that must operate in concert to perform physiological functions. Efforts to create high-dimensional biomarker catalogs of these cells are largely based on transcriptomic single-cell approaches that lack the spatial context required to understand critical cellular communication and correlated structural organization. To probe *in situ* biology with sufficient coverage depth, several multiplexed protein imaging methods have recently been developed. Though these antibody-based technologies differ in strategy and mode of immunolabeling and detection tags, they commonly utilize antibodies directed against protein biomarkers to provide detailed spatial and functional maps of complex tissues. As these promising antibody-based multiplexing approaches become more widely adopted, new frameworks and considerations are critical for training future users, generating molecular tools, validating antibody panels, and harmonizing datasets. In this perspective, we provide essential resources, key considerations for obtaining robust and reproducible imaging data, and specialized knowledge from domain experts and technology developers.

**Introduction**

Mammalian tissues are comprised of a diverse array of cells that possess unique functional attributes and activation states. Many existing methods capture this heterogeneity and complexity by studying tissues and organ systems at single cell resolution— immunohistochemistry (IHC)[1,2], immunofluorescence (IF)[3-6], transcriptomics[7], mass spectrometry[8-10] and cytometry[11-14] all of these approaches represent a trade-off between spatial information and coverage depth. Flow or droplet based single-cell methods, such as multiparameter flow and mass cytometry[11-14], single-cell RNA-sequencing (scRNA-seq)[15,16], and cellular indexing of transcriptomes and epitopes by sequencing (CITE-seq)[17] can define unique cell subsets with incredible granularity. These technologies have greatly expanded our understanding of cell types and states and offer new ways for multi-parametric stratification of samples/patients while identifying potential targets for clinical research. However, they typically require tissue dissociation and do not provide a spatial context for cell-to-cell interactions present in normal tissues and altered in disease[18-20]. Furthermore, these methods fail to retrieve all cell types due to a combination of factors including, but not limited to, differences in dissociation procedures for individual cell types within a tissue[21], cell loss during sorting, and low sampling resulting from cost or sequencing depth.[22] While recent imaging or sequencing based methods probe the spatial transcriptome at single-cell resolution[23-28], *in situ* protein detection overwhelmingly relies on antibodies. Hence, antibodies have been at the core of several new multiplexing approaches that allow detection of spatial cellular organization and composition of tissues at the protein level (**Table S1**).

These multiplexed imaging technologies enable detailed interrogation and characterization of cell types of interest—lymphocytes, stromal cells, structural markers—in human tissues beyond the spectral limitations of conventional fluorescence microscopy (typically <5 targets). As the number of protein biomarkers, >10-50 parameters detected via multiplexing increases (**Fig. 1a**), the assays become increasingly more complex. Therefore, additional effort is required for multiplexed panel design, antibody validation, and careful data acquisition to avoid artifacts while maintaining reproducibility. Future analyses will require robust workflows that yield high quality images and the computational tools needed to optimally mine these data. Here, we provide a summary of several multiplexed antibody-based imaging approaches and outline strategies for sample and custom reagent preparation, rigorous antibody validation, and multiplex panel building. We then discuss unique challenges surrounding the processing, analysis, and storage of imaging data. Finally, we share a perspective on the future of the field that highlights the utility of these methods and provides practical guidelines for widespread adoption for the methods discussed here.

**Multiplexed antibody-based imaging methods**

Multiplexed antibody-based imaging methods can be classified based on the mode of antibody tagging (e.g. metal tag, fluorophore, DNA oligonucleotide barcode, or enzyme) and detection modality (e.g. mass spectrometry, spectroscopy, fluorescence, or chromogen deposition), with each approach providing distinct advantages and disadvantages (**Table S1, Fig. 1b**). Detailed descriptions of these methodologies have been discussed elsewhere[18,29,30], thus, this review focuses on the practical aspects of highly multiplexed imaging experiments. Of these antibody-based imaging modalities, fluorescence-based multiplexed imaging is the most established given the wide availability of reagents and imaging systems (**Fig. 1b**). Typically, fluorescence experiments are limited to visualizing 4-7 antigens within a single imaging cycle due to spectral overlap of selected fluorophores and availability of labelled commercial antibodies. Hyperspectral methods enable data collection beyond this limit, achieving single pass multiplexed imaging of up to 21 channels by utilizing fluorophores with diverse excitation and emission spectra, advanced instruments, and methods that enable compensation for spectral overlap[22,31].

Even higher dimensional datasets can be obtained through an iterative, multi-step process (or cycle) that includes (**Fig. 1b**): 1) immunolabeling with fluorescent or oligonucleotide barcoded antibodies, 2) direct image acquisition (fluorescent antibody-stained material), or reaction with fluorescent complementary oligonucleotides for a subset of the tagged antibodies, followed by image acquisition, and 3) fluorophore inactivation or removal of antibodies or hybridized oligonucleotide probes. This process circumvents spectral overlap by removing fluorescent signals after each cycle. Using iterative staining, imaging, bleaching/antibody removal methods such as tissue-based cyclic immunofluorescence (t-CyCIF), iterative indirect immunofluorescence imaging (4i), iterative bleaching extends multiplexity (IBEX), and multiplexed immunofluorescence (MxIF), it is possible to detect >60 targets in the same tissue section using off-the-shelf antibodies with fluorescent secondaries or fluorescently conjugated primaries[32-40]. A critical consideration for the implementation of such methods is the potential for epitope loss, tissue degradation, and incomplete fluorophore inactivation over successive cycles. For these reasons, it is important to include appropriate controls as outlined here and described previously[33-35,41,42].

Alternative methods that rely on DNA-barcoding of antibodies, such as DNA Exchange Imaging (DEI)[43], immunostaining with signal amplification by exchange reaction (immuno-SABER)[44], and co-detection by indexing (CODEX)[19,45,46], allow one-step immunostaining and fast sequential barcode readout through rapid binding/unbinding of fluorescent oligos. Although these methods can detect a high number of target molecules in a wide range of tissues, their utility is limited in samples where harsh fixation has reduced epitope retention or tissues with high autofluorescence. Amplification methods such as Immuno-SABER[44], immunosignal hybridization chain reaction[47,48], and enzymatic (e.g. horse radish peroxidase[49] or tyramide signal amplification (TSA)[50] approaches can be used to improve signal-to-noise while increasing the detection of low abundance epitopes. Until recently, TSA methods such as Opal IHC[50-52] have been restricted to the simultaneous detection of 6-8 different markers[18] as tyramide-linked fluorophores remain bound to the tissue despite rounds of antibody removal and amplification. Nonetheless, it was recently shown that the fluorophore limitations of Opal IHC could be extended using lithium borohydride to eliminate signal from several Opal dyes, providing a means for the capture of highly multiplexed images in heavily fixed tissues[32].

An alternative to fluorescence-based approaches is mass spectrometry (MS)-based methods that incorporate ionizable metal mass tags (**Fig. 1b**). These technologies enable visualization of >40 biomarkers in a tissue section using a single master mix of metal-conjugated primary antibodies and do not require antibody staining/removal cycles. The two most common technologies are multiplex ion beam imaging (MIBI)[53] and imaging mass cytometry (IMC)[54], which differ by the use of an ion beam or laser, respectively, for tag ionization. The major benefit of these MS-based technologies is the ability to detect and resolve dozens of metal isotope-labelled antibodies simultaneously, aptly named "all-in-one". The metal ion barcodes possess low background signal by circumventing autofluorescence and incorporating high instrumental mass resolving power. Unlike other multiplexed approaches, samples must be vacuum stable and antibody labels cannot be amplified. However, the limit of detection of MS systems is on the order of attomoles. The availability of bovine serum albumin (BSA)-free antibody formulations is a limitation shared with other imaging workflows requiring custom reagent creation, e.g. chelation or fluorophore conjugation. An additional potential barrier associated with expanding metal-tagged antibody panels is access to sufficient amounts of isotopically pure lanthanide metals. While image acquisition occurs in one cycle, it can be slow when considering large fields of view (~200 px/s or 1 mm$^2$/ 2 h)[55]. Finally, to date, the instruments are less operationally stable than conventional fluorescence microscopes and require more specialized staff and environments for effective use.

Other spectroscopy methods that also employ "all-in-one" data collection include multiplexed vibrational imaging[56] using techniques such as stimulated Raman scattering (SRS) or surface enhanced Raman spectroscopy (SERS). Instead of mass barcodes, these methods utilize the enhanced vibrational signatures of fluorophores at different wavelengths for multiplexing up to 22 targets. Sample preparation for spectroscopic multiplexed imaging is similar to fluorescence microscopy and can be performed on various tissue preparations without a vacuum system.

Beyond investigating tissues in 2D, recent advances in sample preparation and imaging have enabled exploration of entire tissue volumes (3D) to allow a deeper understanding of total organ architecture while permitting detailed characterizations of rare cells that are frequently undersampled in thin (5-10 μm) tissue sections[57]. For instance, chemical clearing methods can be applied to transform intact tissues and organs into robust, transparent samples for future antibody staining and imaging. Importantly, clearing methods must preserve the spatial distribution of proteins and nucleic acids while remaining permeable to fluorescent probes. Several tissue clearing options can be used, such as: clear lipid-exchanged anatomically rigid imaging/immunostaining-compatible tissue hydrogel (CLARITY), clearing-enhanced 3D (Ce3D), system-wide control of interaction time and kinetics of chemicals (SWITCH), magnified analysis of proteome (MAP), stabilization under harsh conditions via intramolecular epoxide linkages to prevent degradation (SHIELD), entangled link-augmented stretchable tissue-hydrogel (ELAST), and protein retention expansion microscopy (pro-ExM)[58-64]. After clearing, tissues have been reported to be amenable to antibody staining and subsequent probe removal using either heat, detergents[58,60-62], oligo-probe dehybridization, or fluorochrome reduction. This allows the potential for higher multiplexed 3D imaging. Due to the volume of tissue being analyzed, staining and imaging steps are significantly longer than in conventional histology, although the use of electrophoresis-accelerated antibody transport[65,66] and light-sheet microscopy enable whole murine organs to be stained within a day and imaged within hours. Together, these approaches enable rapid and multiplexed 3D interrogation of intact tissues, providing system-scale structural and molecular information.

**Initial experimental considerations**

We suggest the mantra: "begin with the end in mind" when selecting a multiplexed antibody-based workflow to adopt or implement. The number of samples, whether multiplexed imaging will be a core focus of the outlined work, available budget, and analytical support are key considerations when choosing an approach for spatial profiling of tissues (**Fig. 2a**). As discussed above, each multiplexed imaging method has distinct advantages and disadvantages. Therefore, questions about final data requirements, available samples and format, and existing infrastructure must guide the choice of multiplexed imaging method (**Fig. 2b, Table S1**). Similarly, the first and most essential step of creating a multiplexed imaging panel is to determine the scientific questions to be addressed. As all multiplexed imaging modalities share the process of creating and validating multiplexed antibody panels, we focus the majority of our recommendations on creating a validated panel of 10-60 antibodies (**Fig. 3**).

*Tissue Handling*

The multiplexed antibody-based imaging methods described here are compatible with a wide range of tissue and sample preparations including fresh frozen (FF), formalin-fixed, or formalin-fixed paraffin embedded (FFPE, **Fig. 3**)[19,36]. When deciding on a sample preparation method, one should consider the preservation of tissue architecture, long-term storage conditions, epitope accessibility, and ease of multiplex tissue imaging. To this end, FFPE workflows preserve overall tissue architecture and cellular morphology better than most FF methods[19, 36]. Additionally, most clinical or archival tissues are stored as FFPE blocks due to enhanced preservation and compatibility with room temperature storage. However, FFPE preservation renders many target epitopes inaccessible, increases autofluorescence due to formalin fixation[30,67], and frequently requires antigen retrieval[68,69]. In contrast to single marker IF/IHC, antigen retrieval conditions suitable for one epitope may be incompatible with another epitope, requiring different antigen retrieval methods for distinct molecules, making multiplex panel development increasingly challenging. In contrast, FF tissues overcome the need for antigen retrieval but instead require immediate snap freezing and ultralow temperature storage (<-80 °C) for best tissue preservation[70,71]. Lastly, a tissue preservation method employing 1% paraformaldehyde and a detergent has been shown to preserve tissue architecture, minimize tissue autofluorescence by lysing red blood cells, and enable immunolabeling with a large range of antibodies without the need for antigen retrieval in a variety of human and mouse tissues[32].

*Costs*

An overlooked consideration when designing highly multiplexed imaging experiments is the overall cost and time investment, including reagents, antibodies, training, panel optimization, image acquisition, and data analysis (**Fig. 3**). Ultimately, the cost per unique data point and power of contextual multiplexed data can be useful metrics when rationalizing the increased costs described below. For example, whereas a single-plex experiment can be a few hundred US dollars, a 50-plex panel could cost many thousands of US dollars for an identical number of tissue samples. Several factors contribute to the substantial increase in time and cost of the panel including the complexity of the panel (e.g. number of antibodies, specimen type, sample preparation steps), stability of target epitopes over multiple cycles (if applicable), antibody compatibility and performance under unified immunolabeling conditions, and iterative panel design and adjustment for optimal results. Moreover, identifying and obtaining candidate antibodies can take days to weeks and may additionally require modification of primary antibodies with direct tags. In particular, *in situ* validation can consume weeks or even months for complex panels applied to a new tissue type where each antibody needs to be validated first in single-plex experiments and then assessed for performance within the entire panel. Not all antibody clones are likely to work across all multiplexed technologies, and each multiplexing technology requires sample-specific antibody titration. MS-based platforms are even more expensive, requiring specialized imaging instrumentation and expensive labeling reagents. Beyond raw materials, training requirements vary widely across different assays and methodologies and can dramatically increase costs. The time invested in panel development is higher in fluorescence-based approaches than MS-based approaches because of the cyclic nature of these methods. However, as multiplexed methodologies are commercialized and adopted by core facilities, the time, training, and optimization costs will ultimately decrease.

*Target selection*

Additionally, it is critical to consider downstream analytical requirements: image registration (e.g. nuclear labels such as DAPI or Hoechst 33342)[3-5], cellular segmentation (e.g. robust nuclear and membrane stains)[19,46], and unsupervised clustering (e.g. phenotyping discrete populations using a combination of markers)[68,72-75]. By considering these requirements, putative markers are suggested to be ranked and categorized as "essential" or "desired", greatly expediting overall panel development. This information, along with the budget and approximate timeline, will determine the number of markers to be included in a multiplexed imaging panel.

Marker selection can be guided by expert knowledge, existing literature, orthogonal datasets, and/or online resources (**Table S2**). After establishing a target list of molecules, evaluating and purchasing appropriate antibodies is the next step. In recent years, multiple antibody search engines have been generated, offering intuitive user interfaces that curate antibody clones by citations or artificial intelligence-based approaches (**Table S3**). In addition to identifying highly cited antibody clones, these platforms allow investigators to query many relevant search fields such as tissue and cell type, application, company, and host species while often providing reference images. These resources are a useful starting point for identifying reagents that provide robust immunolabeling in traditional imaging assays. However, these databases do not provide details on performance related to multiplexed imaging, such as epitope stability, steric hindrance, and compatibility with other antibodies. For these reasons, we provide a list of antibody clones previously validated for FF, formalin-fixed, and/or FFPE prepared mouse and human tissues using high content imaging methods to serve as a starting point for creating custom panels (**Supplementary Excel Table**).

For laboratories interested in routine multiplexed imaging, we recommend creating base panels with well-established antibodies that can be expanded with additional antibodies tailored to the questions and tissues of interest. Such an approach reduces the time and capital associated with antibody validation. Moreover, recombinant monoclonal or multiclonal antibodies with appropriate quality control measures, offer the least lot-to-lot variation making them preferable for highly multiplexed panel development. Beyond that, preference is given to rigorously quality controlled hybridoma-derived monoclonal antibodies and, lastly, to polyclonal antibodies. Applications and sample specific testing of each vendor lot is recommended to ensure reproducibility across experiments regardless of antibody type.

**Creating custom reagents**

Many low-plex IHC and IF methods utilize primary antibody detection via a secondary antibody (indirect detection) to enable signal amplification. While convenient, these indirect labeling approaches often cannot be employed to obtain highly multiplexed datasets due to significant overlap between the primary antibody host species and isotype. Consequently, it is often necessary to modify primary antibodies with functional groups that enable detection (**Fig. 4** and **Table S1**). Ultimately, reagent preparation strategies are driven by the cost, availability of materials, technical capabilities, scale, and intended experimental design.

Numerous conjugation methods are used for antibody labeling, each having distinct advantages and disadvantages as well as several tunable variables including, but not limited, to: cross-linker chemistry, cross-linker concentration, pH, modifier concentration, reaction buffer, and purification strategy. Factors like degree of labeling and conjugate purity/homogeneity might vary depending on the method and adjustable variables[76]. Frequently, conjugation chemistries target lysine, cysteine, or glycan residues of the antibody (**Fig. 4**), though there are other site-directed approaches that target the N-terminus[76], defined lysine groups[77,78], or the $F_c$ region of antibodies[79-81]. Furthermore, some non-covalent approaches utilize pre-mixing secondary affinity reagents (e.g. Fab domains, nanobodies or Protein A/G) for multiplexing[82,83].

*Conjugation chemistries*

Lysine-based conjugations[84] are typically preferred for conventional dyes and oligonucleotide barcodes because they are quick, cost effective, controllable, and scalable. A typical IgG has over 40 solvent-exposed lysines, and the extent of conjugation on these residues enables adjustment of the antibody to modifier ratio. While optimizable, the labeling is not site-specific and can result in heterogeneous conjugation and cross-reactivity with non-lysine amino acid residues. If critical lysines are in the antibody active site, the modification can reduce epitope binding activity. These factors can make probe standardization and validation more difficult, although recent publications describe site-specific lysine modification, which may address some of these issues[77,78].

Alternatively, cysteine-based conjugations[85] are also fast, cost effective, scalable, and preferred for the addition of metal tags and oligonucleotide barcodes. While cysteine-based conjugation has been shown to generate more homogeneous and reproducible outcomes, there are only 12 available cysteine sites on a typical IgG. In addition to a lower degree of labeling, the proximity of cysteine residues may result in dyes conjugating near one another, which tends to reduce fluorescence due to quenching. Furthermore, the antibody must be mildly reduced with dithiothreitol or 2-carboxyethyl phosphine (tris) prior to conjugation with maleimide-fluorophore/chelator or cross linkers, which can impact the antibody structure and affinity.

Often, glycan-based approaches are preferred for small scale conjugations, when other protocols fail, or when conjugating oligonucleotide barcodes. On average, an IgG contains two glycan sites on the heavy chain. Because of this, glycan-based conjugations are site-specific, require less optimization of the antibody to modifier ratio, are more reproducible, and avoid binding site inactivation. Unfortunately, the protocol is typically longer and requires enzymatic reactions, which are more costly and difficult to scale.

Standard commercial antibody formulations may not be directly compatible with conjugation chemistries since they often contain sodium azide, bovine serum albumin (BSA), glycerol, and/or cryoprotectants. To facilitate conjugation, we recommend purchasing antibodies in simple buffers such as PBS or performing affinity purification. Further, stocks should be maintained in PBS or borate buffered saline. As each conjugation results in antibody loss, we recommend starting with 50-100 µg of the purified antibody to yield an appreciable product. While conjugation procedures and methods significantly vary, we visually summarize common approaches here **(Fig. 4).**

*Conjugation considerations*

Ultimately, the generation and validation of custom antibody reagents is a major bottleneck for developing multiplexed antibody-based imaging panels. Extended experimental planning and validation of conjugated antibodies associated with multiplexed experiments increases the cost and effort compared to their single-plex counterparts. Custom conjugation also introduces additional batch-to-batch variations, especially for the small-scale preparations done in academic research groups. While commercial custom conjugation services and ready-made conjugation kits make it easier to modify antibodies for custom panels, a wider selection of off-the-shelf primary antibodies conjugated with standard sets of dyes, metals, and oligonucleotide barcode libraries will facilitate their greater adoption. Additionally, antibody performance is frequently affected by the choice of metal tag, fluorophore, or oligonucleotide barcode to which it is conjugated **(Fig. S1)**. For this reason, antibodies that target low abundance epitopes should be conjugated to bright fluorophores that do not overlap with native autofluorescence. Similarly, metal groups and oligonucleotide sequence combinations that have been previously validated for the tissue type are preferred. After custom conjugations, the functionality of the final antibody-conjugate should be confirmed by comparison to the unconjugated version and reassessed with the target assay **(Fig. 4)**. Modified antibodies may also require alterations in tissue blocking and antibody incubation conditions for optimal performance.

**Multiplexed antibody panel design and development**

*Antibody validation*

Antibodies must be validated for each workflow as antibody performance is dependent on tissue type, preservation method, antigen retrieval conditions, final antibody concentration, incubation buffers, and detection method (**Fig. 3**). While best practices for antibody validation have been described by Uhlen *et al.*[86], and have been extensively covered before[42,86-90], for multiplexed antibody-based tissue imaging, we recommend considering: 1) tissue and subcellular location of an antibody target, 2) positive and negative tissue controls, 3) native tissue autofluorescence and other types of background, 4) marker compatibility (e.g. no cross-reactivity or steric hindrance between antibodies), 5) confirming antibody specificity through single-plex IF or IHC experiments[91] and 6) signal-to-noise, particularly for experiments that cannot be amplified (**Fig. S1**). In addition to the antibody-specific search engines discussed here (**Table S3**), regularly updated databases (**Table S2**) can aid in establishing the location and relative abundance of a marker. Using this information, one can validate antibodies using appropriate tissues with well-documented expression of targeted molecules. Additionally, knockout/knockdown cell lines[92,93] or tissues with distinct spatial expression patterns of targets can be used to further validate the specificity of an antibody[86,90]. Finally, pathologists can provide invaluable input into the specificity of antibody staining, particularly in the case of rare clinical samples or atypical markers, where artifactual staining may be more difficult to discern. While we provide an overview of antibody validation for multiplexing here, earlier resources provide a more in-depth and focused discussion of the antibody validation process[30,35,42]. To increase reproducibility and confidence in the results, we recommend including antibody-validation metadata with published work (**Table S4**).

An additional consideration affecting antibody performance is native tissue autofluorescence, which can vary widely across tissue type, disease state, sample preparation, and fixation method[30,67,94]. Autofluorescence can significantly impact the visualization of markers, particularly if these markers are not paired with fluorophores or spectral channels that have higher signal yields. Thus, even if an antibody is validated for its target specificity, its effective use will depend on the anticipated signal in each tissue as compared to tissue autofluorescence, particularly if the label cannot be amplified. Although there is no standard protocol for controlling autofluorescence[95], we provide strategies for minimizing signal from commonly encountered endogenous fluorophores as a resource to the field (**Table S5**).

*Full panel validation*

Many of the multiplexed imaging methods discussed here employ high numbers of antibodies incubated simultaneously or iteratively[18,19,32-36,53,54,96]. Therefore, once the desired marker list is established and antibody specificity validation is complete, it is necessary to validate the full panel as there may be cross-reactivity between antibodies or spectral overlap from reporters/barcodes within the panel that need to be adjusted (**Fig. 3**). When antigen retrieval is required for any of the antibodies in a multiplex panel, the same method of antigen retrieval must work for all antibodies in the panel. Thus, it is crucial to compare specificity of an antibody within a multiplexed panel to its performance within a single-plex experiment. If an antibody does not perform similarly to this single-plex experiment, further adjustments may need to be introduced (e.g. optimization of the staining or conjugation conditions, alternative clones).

Finally, we recommend titrating antibodies over their dynamic range to determine the best signal-to-noise ratio while limiting spectral overlap[18]. Initial antibody validation and multiplexed panel design can be performed in control tissues prepared in the same manner as the experimental tissues. The use of control tissue allows precious samples to be reserved for final image collection using the fully validated panel of antibodies. Cell or tissue microarrays (TMAs) are particularly useful for a quantitative assessment of the optimal antibody titrations on the same slide[87,97]. TMAs, especially multi-organ TMAs, are also recommended for validation of the antibody specificity across multiple tissues in a single staining round. Recently, TMAs were used to validate IMC antibodies and predict biomarkers relevant to disease outcome using AQUA[98-100], a software that uses molecularly defined compartments to automatically measure signal intensity with subcellular resolution. For cyclic methods, it is necessary to verify that tissue morphology and target epitopes remain intact during each stain/bleach cycle or antibody removal steps. Verification methods have been detailed elsewhere and include 1) comparing antibody performance in single-plex (serial) and cyclic (iterative) imaging experiments[32], 2) varying the order of antibody addition (first cycle versus the fourth cycle), and 3) reimaging the same target across multiple cycles[35]. If loss of antibody immunogenicity is observed, we recommend placing the affected antibody earlier within the cycle order, substituting with a different clone, or optimizing the cycling conditions such that low intensity antibodies proceed high intensity antibodies.

**Reproducibility and data analysis**

One of the greatest challenges facing any scientific method involves reproducibility and rigor of the published conclusions[101-103], particularly for antibody-based approaches that are prone to variable results in the absence of tightly controlled experiments[104-106]. Due to the complexity surrounding multiplexed antibody-based experiments, we encourage the inclusion of detailed validation data in published work or deposited as part of publicly available datasets. By mandating and standardizing these processes, we can facilitate appropriate data publishing and use practices, which will undoubtedly improve the experimental rigor and reproducibility. Sharing data on effective antibodies collectively saves an enormous amount of time and resources, but requires extensive, multipronged validation and stable antibody production. We believe that commercial products play a major role in the future adoption and high-fidelity use of multiplexed methods. Access to application-specific vendor validation and quality control data with detailed information (e.g. optimal concentration, tested antigen-retrieval protocols, buffer conditions, stability/storage conditions, RRID/clone information), production of monoclonal antibodies with known sequences/epitopes, broader antibody tag options and flexible formulations (e.g. conjugation compatible buffers, lyophilized products) will greatly support the field. Publicly available data repositories, such as the HuBMAP data portal[107], HTAN data portal[108], The Human Protein Atlas[1], The Cancer Imaging Archive[109], and The Image Data Resource[110], can be leveraged as early domains for sharing and assessing data.

Another large source of variability comes from how and to what extent the raw imaging data are processed. To facilitate this discussion, we propose the use of predefined data states that can describe any multiplexed imaging set (**Table S6**). Many of these processing and segmentation analyses can be performed using a combination of commercial, freeware, and lab-built software packages. Some of the most widely used software packages include Zen (commercial and freeware, Zeiss), ImageJ (open source)[111-114], ASHLAR (open source)[115], cytokit (open source)[116], CellProfiler (open source)[117], CellPose (open source)[118], NAPARI (open source), CODEX Processing and inForm/MAV (commercial, Akoya Biosciences), HALO (commercial, Indica Labs), QUPath (open source)[119], DeepCell (open source)[120], ilastik (open source)[121], Visiopharm TissueAlign/TMA (commercial, Visiopharm), histoCAT (open source)[122], MCMICRO (open source)[123], Squidpy (open source)[124], CytoMAP (open source)[125], and many others[126-128]. Alternative, segmentation-free approaches have also been developed for quantitative analysis of imaging data using pixel-based assignment of subcellular features[33,98,100] as well as classification of irregularly shaped myoepithelial cells[129].

Multiplexed tissue imaging experiments typically result in sizeable files due to multiple imaging cycles for large regions of interest scanned with tiling and z-stacks. To account for this, workstations or computer servers with extensive memory, high processing speeds, efficient graphic cards, integrated storage, and multicore processors are required for handling these large datasets. Generally, processing pipelines work down the data state levels and are, principally, an accumulation of different software packages that transform the raw data into processed and segmented data. Of note, cell type or functional unit annotations typically require incorporating the opinion of a field expert (e.g. pathologist, tissue biologist) which limits the speed and sometimes accuracy of the assignment as such tasks cannot be easily automated.

**Future directions and challenges**

Despite the efforts required for more complicated experimental design and thorough validation, multiplexed imaging experiments are important, and often the only way to visualize diverse cell types and physiological states within complex biological systems *in situ*[130]. For instance, the most recent advances in multiplexed imaging technologies allow over 60 antigens to be stained within a single tissue section; an order of magnitude higher than single pass experiments performed on the best microscopy systems. Most mammalian systems contain hundreds of different cell types that serve essential functions and are subject to regulation by their spatial position and neighbors. Importantly, disease formation, progression, and treatment response may be governed by unique cellular associations within tissues[129,131]. All of these factors necessitate highly multiplexed microscopy approaches to understand the complexity of healthy and diseased tissues. In a recent example, >50 proteins were visualized to reveal how distinct cellular neighborhoods orchestrate the spatial organization of the immune tumor microenvironment and contribute to colorectal cancer outcome[45].

*Extending capabilities and reach*

Future improvements are expected to include improved reproducibility, automated end-to-end workflows, a higher number of markers per experiment, larger imageable areas, and inspiration from outside the field of multiplexed imaging. Of note, a recent study incorporated principles from astronomy to aid in the collection of high-quality datasets from a six-plex TSA-based imaging assay of human melanoma patient samples. The resulting workflow, AstroPath, offers guidelines for operator independent field selection and solutions for commonly encountered problems in multispectral imaging including image processing artifacts, improper image segmentation and classification, and batch-effects due to variation in marker intensities[132]. Importantly, these principles are scalable to the high content imaging assays described here. Another exciting area of development involves computational/automated segmentation of multiplexed images for specific cell types and subtypes to enable tissue-wide assessment of cellular populations. To this end, the HuBMAP consortium[107] has invested significant resources to compare existing cell-segmentation algorithms for their ability to segment membrane and nuclei across multiplexed imaging platforms and defined data processing pipelines. In contrast to the many methods described for cellular segmentation, our ability to segment bulk functional units, such as glomeruli within the kidney or ventricles in the heart, are not as advanced. While individual cells are functionally important, their active coordination within anatomical structures is critical for overall organ function. Multiplexed microscopy approaches are key for advancing this field, as several markers are often needed to define a structure completely and automation enables bulk tissues to be analyzed quickly. Finally, effective implementation of these analytical methods in thicker volumes will greatly advance organ-level studies.

*Dealing with data: standardization, reproducibility, and storage challenges*

While the increased parameter depth is crucial for understanding complex mammalian systems, each additional cycle or protein biomarker provides additional data, often leading to terabytes of raw microscopy images. As such, there are considerable challenges associated with transferring, storing, and sharing raw data. Many journals now require submission of unprocessed data to combat the increasing number of non-reproducible studies that can result in retraction[133-135], particularly with regard to antibody-based approaches. These datasets present challenges as multiplexed experiments often require proprietary software, scale with the number of cycles, markers, and samples captured. The sheer size of datasets alone present unique challenges for storage and handling. For instance, the raw data for a 10 cycle CODEX experiment (200 tiles, 11 z-stacks, 20× objective) is >1.5 TB (150 GB/cycle of 3 probes and DAPI). Many databases/data banks will not store data of this size without considerable cost to the authors or journal, discouraging open data sharing. Processing steps such as compression, extended depth of field, and overlap cropping will reduce the file size considerably (~290 GB using the above example), but this involves altering the image, preventing others from reprocessing the original data with improved algorithms and software. Moreover, data analysis and processing steps should be transparent, such that analysis-specific metadata or processing standards are reported to enhance the reproducibility and reliability of the final dataset in line with FAIR data reporting standards[136]. HTAN consortium has recently provided detailed guidelines and references for components and a structured database for recording of multiplex tissue imaging metadata (Schapiro, Yapp, Sokolov *et al.*, 2021, in preparation). Datasets are most valuable if available in an open/non-proprietary format, such as OME-tiff or BDV, which enables free access to datasets in standardized formats[137]. As an alternative, long-term storage or archiving solutions (cheap but slow) can be utilized for raw data (state 0), whereas higher level data states are maintained on more accessible storage solutions (more expensive but easier to access (**Table S6**)).

Challenges in managing these large datasets are further compounded by the different approaches and tools used to acquire and process the data. Remedying this requires standardized, non-proprietary formats and common infrastructures for sharing and depositing data. We see value in adopting the open science culture and nationally/internationally funded and managed large data repository structures such as the HuBMAP data portal and the Cancer Imaging Archive[109]. This endeavor surely requires a consorted effort from all stakeholders including funding bodies, publishers, method developers, biotechnology companies that commercialize these methods into hardware, software, and chemical products, and the scientists who utilize these resources.

*From multiplexed imaging to sequencing*

As multiplexed imaging approaches are integrated with orthogonal modalities such as flow cytometry, CITE-seq, mass spectrometry, transcriptomics, and spectroscopy, it is crucial to bring the resulting data closer to established "-omics" practices. Incorporating the multiplexed antibody recommendations discussed here, will improve the characterization of cell types and potentially lead to the discovery of cell (sub)types. In addition to independent validation of new discoveries, such multi-modal analysis will enable simultaneous measurement of gene expression and protein products for individual cell types and allow a better understanding of the physiological functions and interacting partners of cells in complex tissues[138]. In addition to the multiplexed imaging methods covered here, new techniques such as DBit-Seq[139] and Digital Spatial Profiling[68] offer a spatially-defined collection of nucleotide barcodes and enable multiplexed protein detection using sequencing platforms. While orthogonal, the best practices described here are largely applicable to these hybrid methods, although distinct considerations may apply for subcellular sampling and sequencing-based detection of targets.

**Conclusion**

To create true atlases of tissues for humans and other species, there is a need for methods that provide the highly resolved spatial context largely missing from isolated cell methods or most tissue nucleic acid profiling technologies. In addition, the latter methods cannot assess cell-cell interactions, ongoing signaling events, how these events acutely affect transcription factor localization, or the translational modifications of proteins. Acquiring such information is the realm of protein imaging methods based on the use of highly specific antibodies. Here, we have provided a primer for multiplexed antibody-based imaging, collated key resources, and outlined considerations for planning and executing such experiments. By following the suggestions presented here and in cited resources, we aim to empower other investigators to obtain high quality imaging data using state-of-the-art multiplexed methods. We believe this primer will result in important contributions to biomedical research and that the emerging findings will be of great impact to the larger scientific community.

**Acknowledgements**

We are grateful for engaging and thoughtful discussions from the Affinity Reagent Imaging and Validation Working Group, HuBMAP Consortium. The authors would like to acknowledge funding from the following sources: NIH U54 DK120058, NIH U54 EY032442, NIH R01 AI145992, NIH R01 AI138581 (R.M.C. and J.M.S.), NIH T32ES007028 (E.K.N), NIH U54 HG010426-01 (M.P.S. and G.P.N), NIH UG3 HL145600-01, NIH UH3 CA246633-01 (R.M.A), NIH UH3 CA246635-01 (N.L.K.), Swedish Research Council 2018-06461 (E.L.), Erling Persson Family Foundation (E.L.), Wallenberg Foundation (E.L.), NIH UH3 CA246594-01 (A.S.), NIH T32CA196585 and ACS PF-20-032-01-CSM (J.W.H.), and NIH UH3 CA255133-03 and European Molecular Biology Laboratory (S.K.S.). This work was supported, in part, by the Intramural Research Program of the NIH, NIAID.

Figures were made using the tools on Biorender.com.

**Author Contributions**

J.W.H., E.K.N., A.J.R, J.M.C., and S.K.S. conceived the plan, designed figures and wrote the manuscript. R.T.B designed display items and helped write the manuscript. A.A., K.C. E.M, J.H., A.E.W., J.F., J.C., A.S., R.M.C., R.M.A., G.P.N., K.C., S.M.H, R.N.G, J.M.S., E.L, M.P.S., and N.L.K provided domain expertise and assisted with the conception and writing of the manuscript.

**Competing Interests**

**Supplementary Information**

Figure S1: Examples for the validation and selection of antibodies for multiplexed antibody-based imaging experiments

Table S1: Summary of multiplexed antibody-based imaging technologies

Table S2: Gene and protein databases for marker discovery efforts

Table S3: Comparison of antibody search engines

Table S4: Suggested minimum antibody metadata needed for multiplexed antibody-based assays

Table S5: Strategies for controlling autofluorescence in mammalian tissues

Table S6: Multiplexed imaging data states

Supplementary Excel Table: Human and mouse antibody clones validated across different highly multiplexed imaging platforms

# Figures

**a**

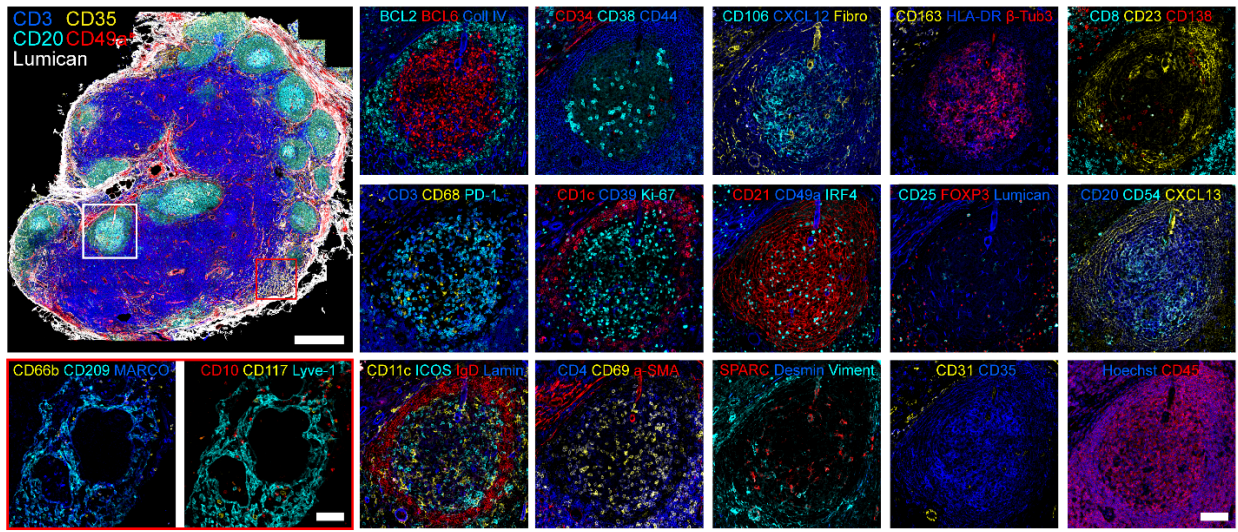

**b**

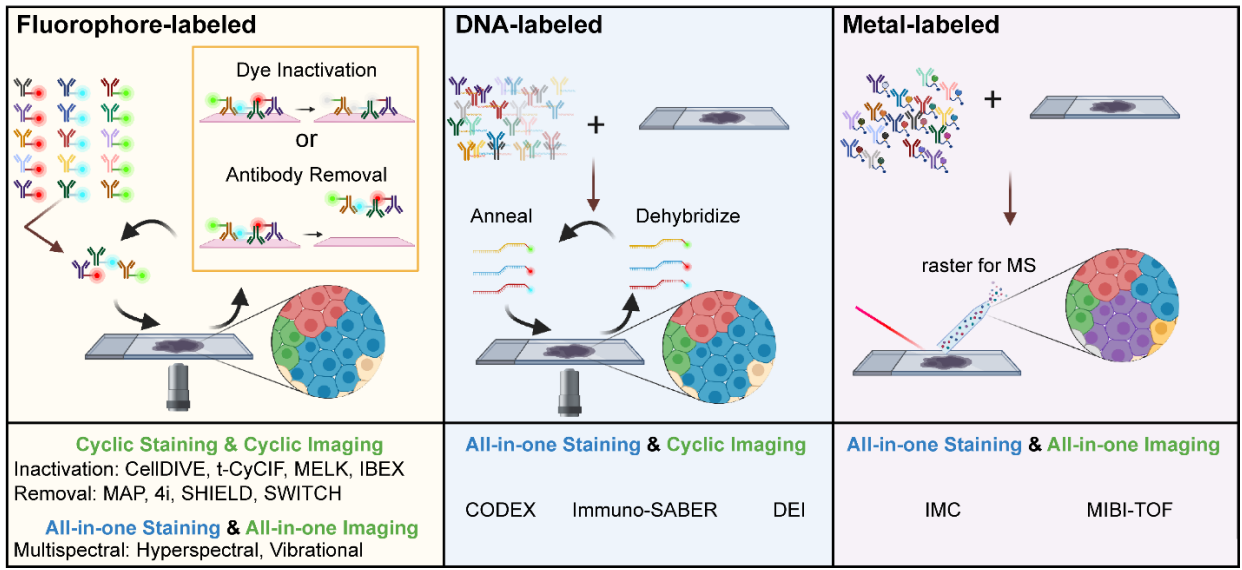

**Figure 1. Obtaining high content imaging data using a wide range of multiplexed antibody-based imaging platforms. a)** Confocal images from a human mesenteric lymph node obtained by IBEX method. Scale bars (500 or 100 μm). β-Tubulin 3 (β-Tub3), collagen IV (Coll IV), fibronectin (Fibro), laminin (Lamin), and vimentin (Viment) (original lymph node dataset from Radtke et al., 2020[32]). **b)** Graphical representation of the main approaches for multiplexed antibody-based imaging. Antibodies are commonly labeled with metals, fluorophores, or DNA oligonucleotides for complementary binding of fluorescent-tagged DNA probes.

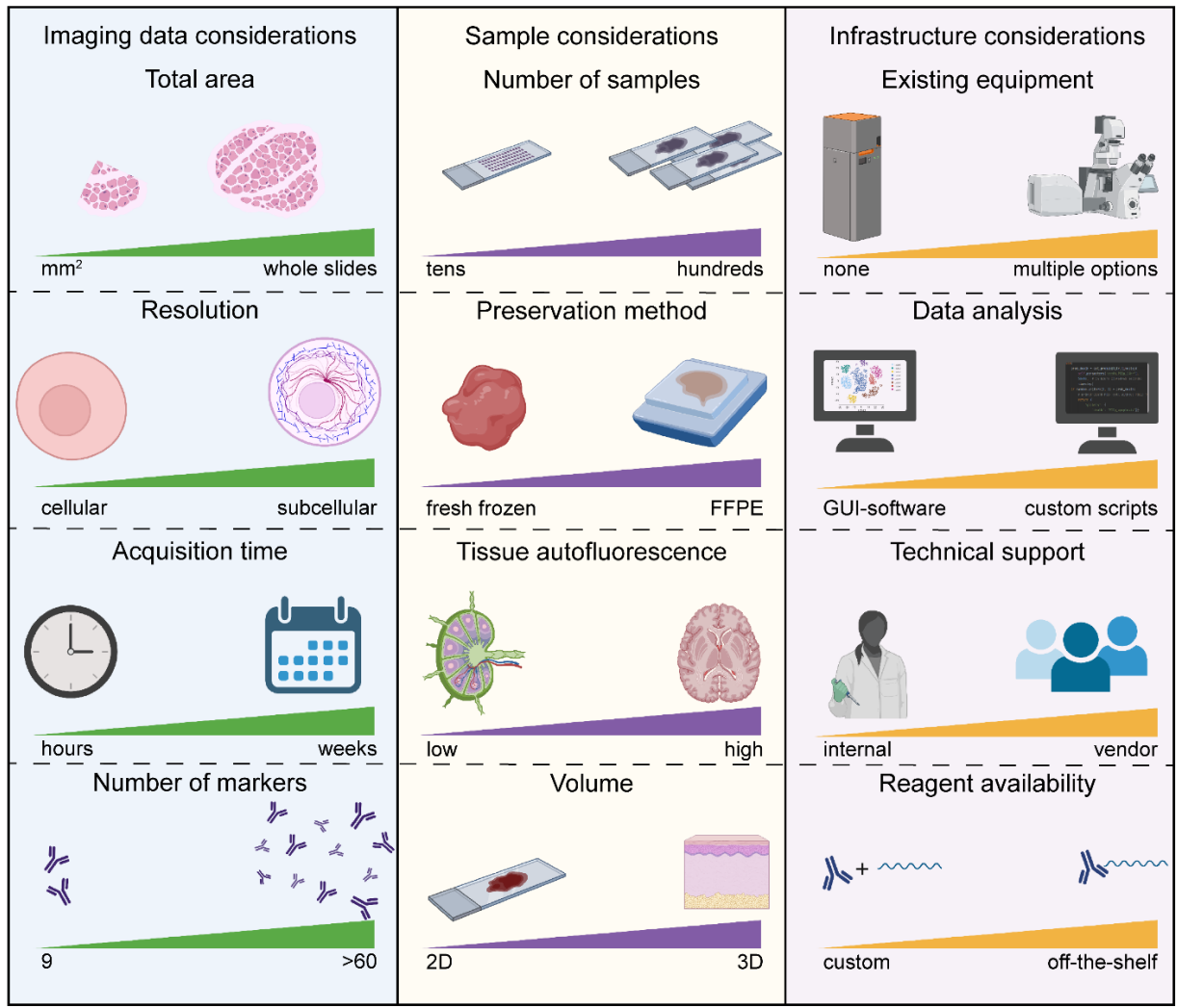

**Figure 2. Considerations for the choice and implementation of multiplexed antibody-based imaging technologies into existing workflows. a)** Open-source, commercial, and core facility options can be separated by advantages/disadvantages related to ease of implementation, initial cost investment, cost per experiment, flexibility and customization, and expertise required. **b)** Several factors govern which method to implement: imaging data requirements (area of tissue needed to be imaged per sample, resolution of final images, time required for imaging each sample, number of markers), sample requirements (number and format of samples, preservation method used for samples, tissue autofluorescence, and whether 2D or 3D volume data are needed), and infrastructure requirements (where existing equipment can be leveraged, level of bioinformatics needed for analysis, technical support, and whether reagents can be purchased or must be customized). Comparisons between several multiplexed imaging techniques are summarized in **Table S1** and described in greater detail in other reviews[18,29,30].

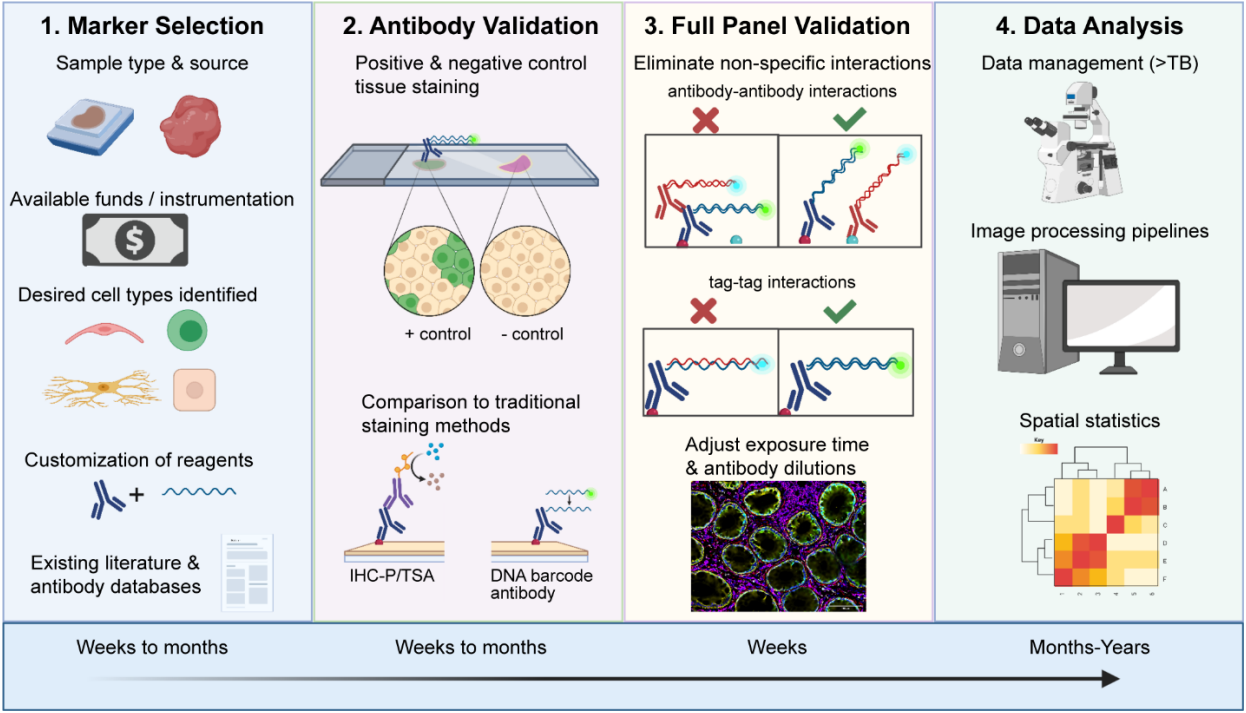

**Figure 3. Phases of panel development and validation for multiplexed antibody-based imaging assays.** Graphical representation of assay development. In Step 1, markers are selected based on indicated criteria. Some multiplexing methods require custom reagents (i.e. directly conjugated primary antibodies) see Figure 4 for more details. In Step 2, antibodies are validated individually to verify target specificity. In Step 3, the full panel is validated to ensure that inclusion of additional antibodies does not affect target specificity. In Step 4, data are collected and analyzed.

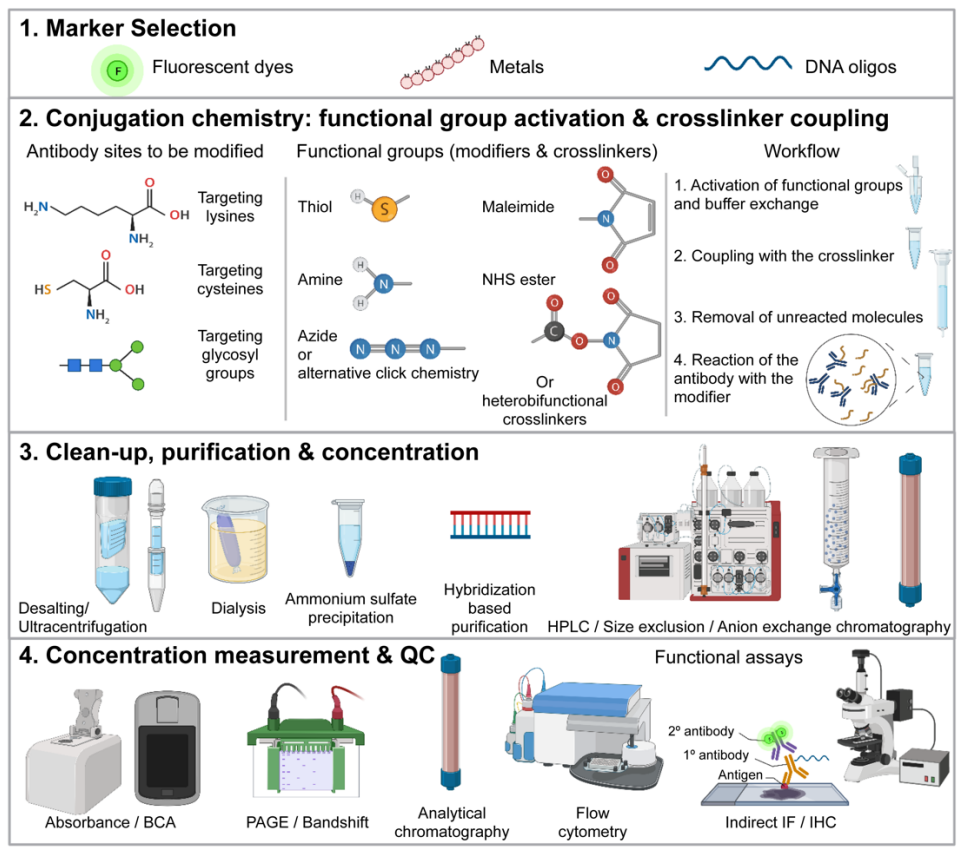

**Figure 4. Process of conjugating antibodies with modifiers for multiplexing.** Following the selection of labeling chemistry and crosslinker, the workflow typically consists of these steps: (1) activation of modifier groups (for example reducing of thiol groups) and buffer exchange, (2) reaction with the crosslinker, (3) removal of the excess crosslinker, and (4) reaction of the antibody with the modifier. Since the reaction typically includes an excess of the modifier, the unreacted molecules can be optionally removed from the mixture using buffer exchange or gel filtration or sequence-directed pull-downs in the case of oligonucleotide barcode modifiers. Further purification can be performed by ion exchange or size exclusion chromatography for removal of aggregated or degraded molecules, leftover modifiers, and under or over-conjugated antibodies. While purified reagents often improve quality and performance, purification may substantially reduce conjugated antibody yield, making it impractical or costly for small scale conjugations. The final product and its purity can be validated by a gel shift on SDS-PAGE gels where both the antibody and the modifier can be observed (for example using Coomassie, fluorescence, Sybr stains, or silver stain). The resulting antibody concentration can be determined using bicinchoninic acid assay (BCA) or absorbance measurements (e.g. NanoDrop), although certain crosslinkers may interfere with these absorbance measurements and corrections need to be made accordingly. Crucially, final product should be re-validated for the general target binding specificity (for example by flow cytometry) or directly for the assay of interest by functional comparison to the unconjugated antibody using direct detection or secondary antibody detection.

# Spatial mapping of protein composition and tissue organization: a primer for multiplexed antibody-based imaging

John W. Hickey[1‡], Elizabeth K. Neumann[2,3‡], Andrea J. Radtke[4‡*], Jeannie M. Camarillo[5], Rebecca T. Beuschel[4], Alexandre Albanese[6,7,¥], Elizabeth McDonough[8], Julia Hatler[9], Anne E. Wiblin[10], Jeremy Fisher[11], Josh Croteau[12], Eliza C. Small[13], Anup Sood[8], Richard M. Caprioli[2,3,14], R. Michael Angelo[1], Garry P. Nolan[1], Kwanghun Chung[6,7,15-18], Stephen M. Hewitt[19], Ronald N. Germain[4], Jeffrey M. Spraggins[3,14,20], Emma Lundberg[21], Michael P. Snyder[22], Neil L. Kelleher[5], Sinem K. Saka[23,24*]

[1]Department of Pathology, Stanford University School of Medicine, Stanford, CA 94305, USA.
[2]Department of Biochemistry, Vanderbilt University, Nashville, TN, 37232 USA.
[3]Mass Spectrometry Research Center, Vanderbilt University, Nashville, TN 37232, USA.
[4]Lymphocyte Biology Section and Center for Advanced Tissue Imaging, Laboratory of Immune System Biology, NIAID, NIH, Bethesda, MD 20892, USA.
[5]Department of Chemistry, Molecular Biosciences and the National Resource for Translational and Developmental Proteomics, Northwestern University, Evanston, IL 60208, USA.
[6]Institute for Medical Engineering and Science, MIT, Cambridge, MA, USA.
[7]Picower Institute for Learning and Memory, MIT, Cambridge, MA, USA.
[8]GE Research, Niskayuna, NY,12309, USA.
[9]Antibody Development Department, Bio-techne, Minneapolis, MN 55413, USA.
[10]Department of Research and Development, Abcam PLC, Discovery Drive, Cambridge Biomedical Campus, Cambridge, CB2 0AX, UK.
[11]Department of Research and Development, Cell Signaling Technology, Inc., Danvers, MA 01923, USA.
[12]Department of Applications Science, BioLegend, San Diego, CA 92121, USA.
[13]Thermo Fisher Scientific, Rockford, IL, USA.
[14]Department of Chemistry, Vanderbilt University, Nashville, TN 37232, USA.
[15]Department of Chemical Engineering, MIT, Cambridge, MA, USA.
[16]Department of Brain and Cognitive Sciences, MIT, Cambridge, MA, USA.
[17]Center for Nanomedicine, Institute for Basic Science (IBS), Seoul, Republic of Korea.
[18]Yonsei-IBS Institute, Yonsei University, Seoul, Republic of Korea.
[19]Laboratory of Pathology, Center for Cancer Research, National Cancer Institute, National Institutes of Health, Bethesda MD 20892, USA.
[20]Department of Cell and Developmental Biology, Vanderbilt University School of Medicine, Nashville, TN 37240, USA.
[21]Science for Life Laboratory, School of Engineering Sciences in Chemistry, Biotechnology and Health, KTH–Royal Institute of Technology, Stockholm, Sweden.
[22]Department of Genetics, Stanford University School of Medicine, Stanford, California 94305, USA.
[23]Wyss Institute for Biologically Inspired Engineering at Harvard University, Boston, MA 02115, USA.
[24]European Molecular Biology Laboratory (EMBL), Genome Biology Unit, 69117 Heidelberg, Germany.
‡Authors contributed equally to this work and are listed alphabetically.
¥Current Address: Boston Children's Hospital, Division of Hematology/Oncology, Boston, MA, USA
*Corresponding Authors

**Contents**

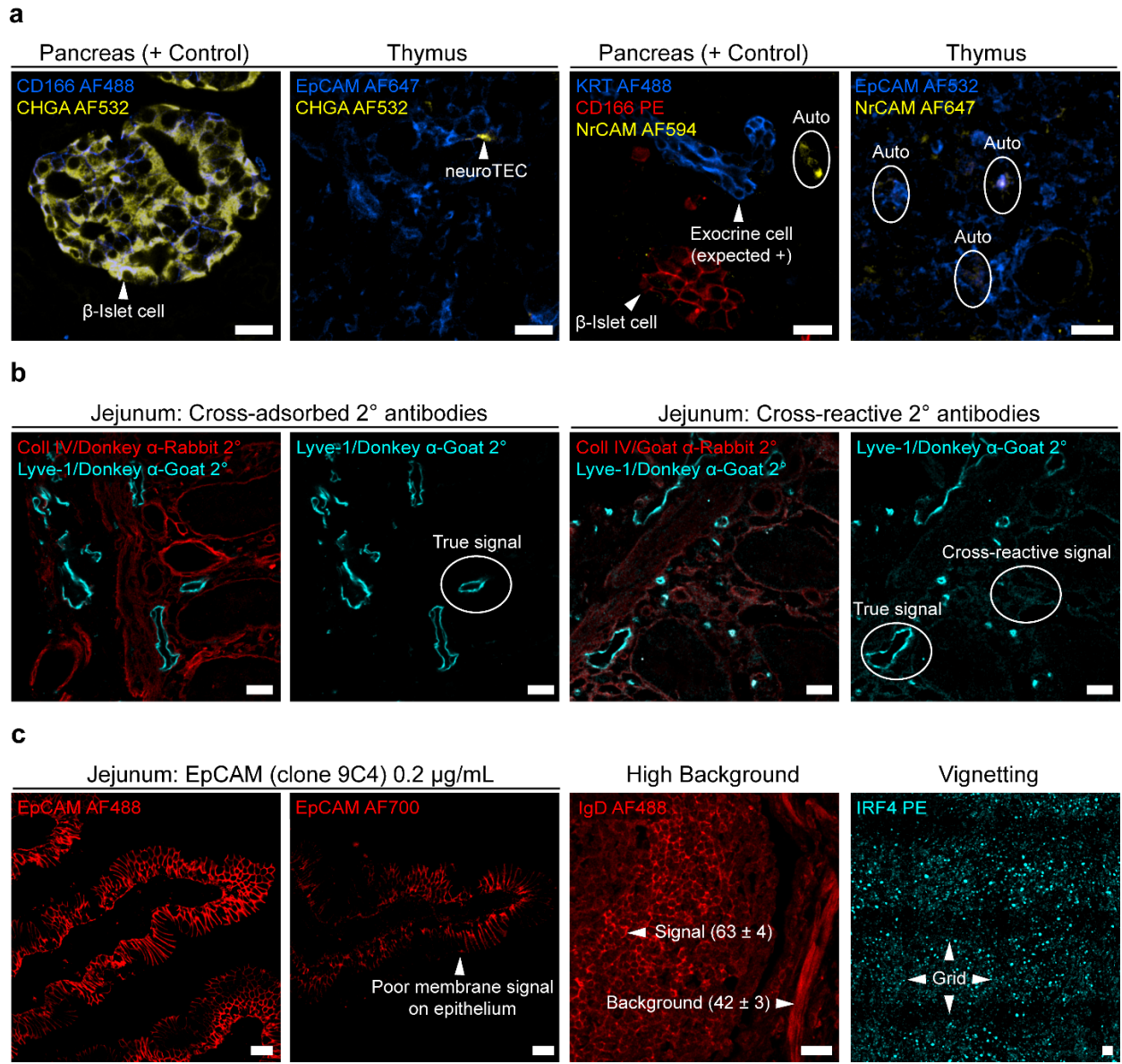

**Figure S1. Examples for the validation and selection of antibodies for multiplexed antibody-based imaging experiments. a)** Confocal images from human pancreas (positive control) and thymus immunolabeled with the indicated antibodies, chromogranin A (CHGA), pan-cytokeratin (KRT), and neuronal cell adhesion molecule (NrCAM). Alexa Fluor (AF) denotes fluorescent conjugate used here and throughout. CHGA positively stains the β-islet cells of the pancreas as well as rare thymic epithelial cells (neuroTEC), demonstrating an appropriately validated antibody. In contrast, the NrCAM antibody does not colocalize with exocrine cells in the pancreas and yields non-specific immunolabeling (autofluorescence) in the thymus. **b)** Confocal images from human jejunum demonstrating the need for cross-adsorbed host-matched secondary (2°) antibodies. Representative true vs. cross-reactive signals are highlighted. **c)** Confocal images from human jejunum and lymph node (right-most panels) demonstrating the impact of antibody conjugate on the fidelity of immunolabeling and image quality. Mean ± SEM pixel intensity for signal and background denoted in parantheses. Scale bars correspond to 25 μm.

**Table S1. Summary of multiplexed antibody-based imaging technologies.** See Taube *et al.*, 2020[1] and Tan *et al.*, 2020[2] for additional comparisons, e.g. cost, scanning area, and total imaging time. Fixed*: 1-4% paraformaldehyde (PFA) or acetone-fixed cryosections. FF: Fresh frozen.

| Method | Antibody Staining Cyclic vs All-in-one | Probe Detection Cyclic vs All-in-one | Antibody Conjugate | Marker Removal Method | Sample Type (Reported) | Maximum Number of Markers (Reported) | Key Reference |
|---|---|---|---|---|---|---|---|
| 4i[3] | Cyclic | All-in-One | Fluorescent | Removal | Cells | >40 | DOI: 10.1126/science.aar7042 |
| MELK[4] | Cyclic | All-in-One | Fluorescent | Dye Inactivation | Cells, Fixed* | 100 | DOI: 10.1007/3-540-36459-5_8 |
| t-CyCIF[5] | Cyclic | All-in-One | Fluorescent | Dye Inactivation | Cells, FF, FFPE | 60 | DOI: 10.7554/eLife.31657 |
| MxIF, CellDIVE[6] | Cyclic | All-in-One | Fluorescent | Dye Inactivation | Cells, FF, FFPE | >60 | DOI: 10.1073/pnas.1300136110 |
| Multiplex IHC[7] | Cyclic | Cyclic + Amplification | Chromogenic | Removal | FFPE | 12 | DOI: 10.1016/j.celrep.2017.03.037 |
| Opal IHC[8] | Cyclic + Amplification | All-in-One | Fluorescent or Chromogenic | Removal | FF, FFPE | 4-7 | DOI: 10.1016/j.ymeth.2014.08.016 |
| IBEX[9] | Cyclic | All-in-One | Fluorescent | Dye Inactivation | Fixed*, FFPE | >60 | DOI: 10.1073/pnas.2018488117 |
| Ce3D[10] | Cyclic | All-in-One | Fluorescent | NA | Fixed | 5-7 | DOI: 10.1073/pnas.1708981114 |
| SWITCH[11] | Cyclic | All-in-One | Fluorescent | Removal | FF, Fixed* | >20 | DOI: 10.1016/j.cell.2015.11.025 |
| SHIELD[12] | Cyclic | All-in-One | Fluorescent | Removal | FF, Fixed* | 10 | DOI: 10.1038/nbt.4281 |
| MAP[13] | Cyclic | All-in-One | Fluorescent | Removal | FF, Fixed* | 20 | DOI: 10.1038/nbt.3641 |
| CLARITY[14] | Cyclic | All-in-One | Fluorescent | Removal | FF, Fixed* | 10 | DOI: 10.1038/nature12107 |
| CODEX[15] | All-in-One | Cyclic | DNA-barcode | Removal | Cells, FF, Fixed*, FFPE | 60 | DOI: 10.1002/eji.202048891 |
| Immuno-SABER[16] | All-in-One | Cyclic + Amplification | DNA-barcode | Removal | Cells, Fixed*, FFPE | >10 | DOI: 10.1038/s41587-019-0207-y |
| DEI[17], Exchange-PAINT[18] | All-in-One | Cyclic | DNA-barcode | Removal | Cells, Fixed* | >10 | DOI: 10.1021/acs.nanolett.7b02716 DOI: 10.1039/C6SC05420J |
| MIBI-TOF[19] | All-in-One | Cyclic | Metal | NA | Cells, FF, FFPE | >40 | DOI: 10.1038/nm.3488 |
| IMC[20] | All-in-One | All-in-One | Metal | NA | Cells, FF, FFPE | >40 | DOI: 10.1038/nmeth.2869 |
| Hyperspectral[21] | All-in-One | All-in-One | Fluorescent | NA | Fixed | 8-21 | DOI: 10.1016/j.immuni.2012.07.011 Orion RareCyte (https://rarecyte.com/orion/) |
| Vibrational[22] | All-in-One | All-in-One | Fluorescent | NA | Cells, Fixed | >20 | DOI: 10.1038/nature22051 |

**Table S2. Gene and protein databases for marker discovery efforts.** Comparison of major online gene and protein databases based on several criteria. *Subcellular location information is predicted and level of information varies for each protein/target of interest; **Does not provide exhaustive visual examples, limited to the most common tissues; ***Based on the UniProt information; ****Limited number of tissues analyzed.

| Database | Type of Information | | | | | | Unique Feature | Links |
|---|---|---|---|---|---|---|---|---|
| | Gene | Protein | Tissue Location | Cell Type Expression | Subcellular Localization | Pathology | | |
| Human Protein Atlas[23] | ● | ● | ● | ● | ●* | ● | Images of markers within tissues | https://www.proteinatlas.org |
| UniProt[24] | ● | ● | ●** | - | ● | ● | Basis for most structural information | https://www.uniprot.org |
| National Center for Biotechnology Information (NCBI)[25] | ● | ● | ● | - | ● | ● | Hosts multiple databases | https://www.ncbi.nlm.nih.gov |
| GeneCards[26] | ● | ● | - | - | ●*** | ● | Collates information from multiple databases | https://www.genecards.org |
| Cell Marker[27] | ● | ● | ● | ● | - | - | Tissue-specific biomarkers | http://bio-bigdata.hrbmu.edu.cn/CellMarker |
| Atlas of Genetics and Cytogenetics in Oncology and Haematology[28] | ● | ●*** | ● | - | - | ● | Cancer focused | http://atlasgeneticsoncology.org |
| Expression Atlas[29] | ● | - | ● | ● | - | ● | Expression across species and conditions | https://www.ebi.ac.uk/gxa/home |
| The NCI Genomic Data Commons[30] | ● | - | ● | ● | ● | ● | Cancer focused with information on mutations | https://gdc.cancer.gov |
| The Human Proteome Map[31] | - | ● | ●**** | ● | ● | - | NA | http://www.humanproteomemap.org |
| ASCT+B Reporter[32] | ● | ● | ● | ● | - | - | Gene and protein biomarkers by cell type and organ, State-of-the-art visualization tool | https://hubmapconsortium.github.io/ccf-asct-reporter/ |

**Table S3. Comparison of antibody search engines.** Comparison of major online antibody search engines. This table is adapted with permission from Dr. Maurice Shen on BenchSci (https://blog.benchsci.com/antibody-search-engines). *Does not list publications from closed-access journals.

| Search Engine | No. of Antibodies (Relative) | Publications Listed/Used in Literature | Ranked by Citations | Independent Validation | Experimental Applications Listed | User Reviews Provided | Links |
|---|---|---|---|---|---|---|---|
| Antibodies.com | 50,000+ | ● | ● | - | - | ● | https://www.antibodies.com |
| Antibodies-online | 831,000+ | ● | - | ● | - | - | https://www.antibodies-online.com |
| Antibodypedia[33] | 4.3 million | ●* | - | ● | - | - | https://www.antibodypedia.com |
| Antibody Registry | 2.4 million | - | - | ● | - | - | https://www.antibodyregistry.org |
| BenchSci | 6.9 million | ● | ● | ● | ● | - | https://www.benchsci.com |
| BioCompare | 3 million | ● | - | - | ● | ● | https://www.biocompare.com |
| BIOZOL | 3.1 million | - | - | - | - | - | https://www.biozol.de/de |
| CiteAb[34] | 5.1 million | ●* | ● | - | ● | - | https://www.citeab.com |
| IHC World | 18,000+ | - | - | - | - | ● | http://www.ihcworld.com |
| Labome | 399,000+ | ●* | - | ● | - | ● | https://www.labome.com/index.html |
| Linscott's Directory | 2.6 million | - | - | ● | - | - | https://www.linscottsdirectory.com |

**Table S4. Suggested minimum antibody metadata to be reported for multiplexed antibody-based assays.** Proposed metadata to be collected on individual antibodies implemented in multiplexed antibody-based imaging assays. Validation should be performed according to previously published recommendations[35-39]. Also further see *Schapiro, Yapp, Sokolov et al., 2021*, for guidelines on a detailed metadata database structure encompassing all steps of multiplex experiments.

| Suggested minimum antibody metadata | Rationale |
| --- | --- |
| UniProt Accession Number | Identifies the target protein. |
| Target Name | Provides a common name for the target protein. |
| RRID | Allows for universal identification of an antibody. |
| Antibody Name | Provides common name for the antibody used. |
| Host Organism and Isotype | Describes the species in which the antibody was raised (e.g. Mouse IgG1). |
| Clonality | Identifies the antibody as monoclonal or polyclonal. |
| Vendor | Provides information on the source of the antibody. |
| Catalog Number | Provides information on the source of the antibody. |
| Lot Number | Allows for monitoring of lot-to-lot variation. |
| Recombinant | Classifies the antibody as recombinant or not. |
| Dilution/Concentration | Provides a recommended usage (e.g. 1:100, 5 µg/ml). |
| Conjugate/Format | Offers details on the format and mode of detection (e.g. Fluorophore (AF647), Metal isotope (164Er), Oligo barcode and detection sequences. If custom conjugated, identify conjugation kit (Vendor, Catalog Number) or conjugation chemistry/protocol. |
| Downstream Platform Used | Identifies the imaging platform used (e.g. CODEX, MIBI, etc.). |
| Organ/tissue/cell line and sample preparation used for validation | Provides details on sample format and preparation protocol (e.g. fixation and antigen retrieval method if applicable). |
| Cycle Number | Identifies order of antibody immunolabeling or visualization for cyclic imaging methods. |
| Special Considerations (If known) | Describes whether a particular antibody is sensitive to imaging order (must go first or last in cyclic imaging methods) or is incompatible with other antibodies (steric hindrance, cross-reactivity with unconjugated antibodies of the same host species). |
| DOI for Validation Protocol | Details the protocol used to validate the antibody including positive and negative controls and example images. To enhance reproducibility, we recommend that all protocols be made publicly available via protocols.io. |
| ORCID ID of author | Identifies the individual who validated the antibody used in the assay. |

**Table S5. Strategies for controlling autofluorescence in mammalian tissues.** Protocols for minimizing the impact of endogenous fluorophores on fluorescence-based multiplexed imaging. Please see Croce *et al.*[40], Davis *et al.*[41], Billinton and Knight[42], Whittington and Wray[43], and (https://hwpi.harvard.edu/files/iccb/files/autofluorescence.pdf?m=1465309329) for additional details.

| Origin[40-44] | Excitation/ Emission (nm)[40,42,44,45] | Tissue Prevalence | Fixative (Note 1) | Quenching (Note 2) | Masking (Note 3) | Light Irradiation (Note 4) | Narrow Ex/Em (Note 5) | Fluorophore Selection (Note 6) | Amplify Signal (Note 7) | Computational (Note 8) |
|---|---|---|---|---|---|---|---|---|---|---|
| Formalin-induced | Broad | Many | +++ | +++ | - | ++ | ++ | ++ | +++ | +++ |
| Bilirubin | (366-465)/ (517-523, 570) nm | Blood, Liver | - | Unknown | Unknown | Unknown | Unknown | Unknown | +++ | +++ |
| Cytokeratins Collagen Elastin | (280-420)/ (400-525) nm | Many, especially, connective tissues, lung, skin, small and large bowel | Unknown | ++ | ++ | +++ | ++ | +++ | +++ | +++ |
| Eosinophils | Broad | Many | - | ++ | ++ | Unknown | - | - | ++ | +++ |
| Flavins | (350-370; 440-450)/ (480/540) nm | Many | + | +++ | Unknown | +++ | ++ | ++ | +++ | +++ |
| Fatty Acids | (330-350)/(470-480) nm | Many, especially adipose, brain, liver, and retina | ++ | +++ | +++ | +++ | - | ++ | +++ | +++ |
| NAD(P)H | (330-380)/ (440, 462, bound, free) nm | Many | + | +++ | Unknown | +++ | ++ | ++ | +++ | +++ |
| Vitamin A | (370-380)/(490-510) nm | Liver, pancreas | Unknown | +++ | Unknown | +++ | ++ | ++ | +++ | +++ |
| Heme Derivatives | (405 nm)/(630-700) nm | Many, especially liver, spleen, RBC-rich tissues | +++ | ++ | +++ | +++ | - | - | ++ | +++ |
| Lipofuscin | Broad | Many, especially adipose, brain, liver, and retina | ++ | +++ | +++ | +++ | - | - | +++ | +++ |

Reduction in autofluorescence: No (-), Minimal (+), Moderate (++), Significant (+++)

**Notes**
1. Avoid fixatives with aldehydes[40] or use a fixative with a detergent to lyse red blood cells and minimize autofluorescence from fatty acids and lipofuscin[9]. Further details on alterative fixatives to reduce autofluorescence from bilirubin[46] and eosinophil granules[47] have been described but are unlikely to be applicable for high content imaging.
2. Chemical or physical quenching with sodium borohydride[40,41], pre-treatment with $H_2O_2$ in the presence of light[5], application of TrueVIEW (https://www.future-science.com/doi/pdf/10.2144/btn-2017-0117), or peroxidase staining reaction with diaminobenzidine (DAB) to reduce autofluorescence from eosinophil granules[47]
3. Masking with dyes such as Sudan Black B[41,48], TrueBlack (Biotium)[43], or a combination of Sudan Black B and 3,3'-diaminobenzidine[49]
4. Light irradiation[40,50,51]
5. Utilizing microscopes with defined excitation sources and narrow filters and/or detector ranges[42,48,52]
6. Selecting fluorophores with high quantum yields[52]
7. Amplifying signal in highly autofluorescent channels with secondary antibodies or TSA-based reagents
8. Computational methods to remove autofluorescence after image acquisition, e.g. spectral unmixing[49], channel thresholding[42], and background subtraction[5,53]

**Table S6. Multiplexed imaging data states.** Table defining common image processing, segmentation, and annotation steps and definitions. We divide each process into unique data states to describe images stored in repositories or submitted alongside manuscripts.

| State | Level of Processing | Definitions |
|---|---|---|
| 0 (Raw Data) | Raw Data | Data that comes directly from the imaging platform |
| 1 (Processed) | Stitching, tiling, thresholding, background subtraction, deconvolution, alignment, and extended depth of field. | **Flat field and dark frame correction:** Corrects for artifacts in illumination and detection<br><br>**Stitching:** Combines multiple images with overlapping fields of view to produce a single image<br><br>**Thresholding:** Identifies a pixel as either positively stained or unstained<br><br>**Background subtraction:** Removes the total intensity signal resulting from autofluorescence or other endogenous output<br><br>**Image alignment:** Aligns each field of view within a defined dimension (e.g., x, y, z, or temporal)<br><br>**Deconvolution:** Computationally compensates for optical distortion to improve image clarity and sharpness<br><br>**Tiling:** Collects individual fields of view for building a larger image using stitching<br><br>**Extended depth of field:** Builds a large, in focus imaging volume by acquiring a z-stack with optimal focus at each position |
| 2 (Segmented) | Nuclear, cellular, and structural. | **Nuclear segmentation:** Automatically partitions pixels into nuclear boundaries<br><br>**Cellular segmentation:** Automatically partitions pixels into cellular boundaries<br><br>**Structural segmentation:** Automatically partitions pixels into structural/functional features |
| 3 (Annotated) | Healthy vs diseased, cell type, functional regions | Interpretation of images by a trained expert, such as a pathologist, physician, or biologist. |

**Supplementary Excel Table**

https://docs.google.com/spreadsheets/d/1PsPnjidPwIUW1jhXtc0254t1nkMns8yJ9sBptsQNxqw/edit?usp=sharing

**Supplementary References**